\documentclass[a4paper,11pt]{article}
\usepackage[left=20mm,right=20mm,top=20mm,bottom=20mm]{geometry}
\usepackage{amsmath}
\usepackage{amsfonts}
\usepackage{amssymb}
\usepackage{amsbsy}
\usepackage{color,graphics}
\usepackage{setspace}
\usepackage{epsfig}
\usepackage{cite}
\usepackage{graphicx}
\usepackage{latexsym,multicol}
\usepackage{appendix}
\newtheorem{theorem}{Theorem}[section]

\newtheorem{proposition}[theorem]{Proposition}

\graphicspath{{./SEC-figs/}}
\begin{document}
\title{Likely cavitation in stochastic elasticity}
\author{L. Angela Mihai\footnote{School of Mathematics, Cardiff University, Senghennydd Road, Cardiff, CF24 4AG, UK, Email: \texttt{MihaiLA@cardiff.ac.uk}}
	\qquad Danielle Fitt\footnote{School of Mathematics, Cardiff University, Senghennydd Road, Cardiff, CF24 4AG, UK, Email: \texttt{FittD@cardiff.ac.uk}}
	\qquad Thomas E. Woolley\footnote{School of Mathematics, Cardiff University, Senghennydd Road, Cardiff, CF24 4AG, UK, Email: \texttt{WoolleyT1@cardiff.ac.uk}}
	\qquad Alain Goriely\footnote{Mathematical Institute, University of Oxford, Woodstock Road, Oxford, OX2 6GG, UK, Email: \texttt{goriely@maths.ox.ac.uk}}
}	\date{October 13, 2018}
\maketitle

\begin{abstract}
We revisit the classic problem of elastic cavitation within the framework of stochastic elasticity. For the deterministic elastic problem, involving homogeneous isotropic incompressible hyperelastic spheres under radially symmetric tension, there is a critical dead-load traction at which cavitation can occur for some materials. In addition to the well-known case of stable cavitation post-bifurcation at the critical dead load, we show the existence of unstable snap cavitation for some isotropic materials satisfying Baker-Ericksen inequalities. For the stochastic problem, we derive the probability distribution of the deformations after bifurcation. In this case, we find that, due to the probabilistic nature of the material parameters, there is always a competition between the stable and unstable states. Therefore, at a critical load, stable or unstable cavitation occurs with a given probability, and there is also a probability that the cavity may form under smaller or greater loads than the expected critical value. We refer to these phenomena as `likely cavitation'. Moreover, we provide examples of homogeneous isotropic incompressible materials exhibiting stable or unstable cavitation together with their stochastic equivalent.  \\
	
\end{abstract}

%%%%%%%%%%%%%%%%%%%%%%%%%%%%%%%%%%%%%%%%%%%%%%%%%%%%%%%%%%%%
%%%%%%%%%%%%%%%%%%%%   NEW SECTION  %%%%%%%%%%%%%%%%%%%%%%%%
%%%%%%%%%%%%%%%%%%%%%%%%%%%%%%%%%%%%%%%%%%%%%%%%%%%%%%%%%%%%
\section{Introduction}

Experiments carried out by Gent and Lindley in 1958 \cite{Gent:1959:GL}, on rubber cylinders, revealed that some materials can rupture under relatively small tensile dead loads by opening an internal cavity. Following this work, the theoretical analysis of Ball (1982) \cite{Ball:1982} provided an explanation for the formation of a spherical cavity at the centre of a sphere of isotropic hyperelastic incompressible material in radially symmetric tension under prescribed surface displacements or dead loads. There, the word `cavitation' was used to describe such void-formation within a solid by analogy to the similar phenomenon observed in fluids. As cavitation in solids is an inherently nonlinear mechanical effect, not captured by the linear elasticity theory, many different studies have been devoted to the modelling of this effect within the finite elasticity framework. For instance: spheres of particular homogeneous isotropic incompressible materials were discussed in \cite{ChouWang:1989:CWH}; homogeneous anisotropic spheres with transverse isotropy about the radial direction were examined in \cite{Antman:1987:AN,Merodio:2006:MS,Polignone:1993a:PH}; concentric homogeneous spheres of different hyperelastic material were analysed in \cite{Horgan:1989:HP,Sivaloganathan:1992,Polignone:1993b:PH}; non-spherical cavities were investigated in \cite{James:1992:JS}; cavities with non-zero pressure were presented in \cite{Sivaloganathan:1999}; cavitation in an elastic membrane was studied in \cite{Steigmann:1992}; the homogenisation problem of nonlinear elastic materials was treated in \cite{LopezPamies:2009,LopezPamies:2011:etal}; growth-induced cavitation in nonlinearly elastic solids was explored in \cite{Goriely:2010:GMV,McMahon:2010:McMG,Pence:2007:PT}. Recent experimental results on the onset, healing and growth of cavities in elastomers were reported in \cite{Poulain:2017:PLLPRC,Poulain:2018:PLPRC}. For many other results on cavitation in solids, we refer to the review articles \cite{Fond:2001,Gent:1991,Horgan:1995:HP}, focusing on rubberlike materials, and the references therein.

The present work focuses on the phenomenon of “cavitation” contained within the theoretical context of finite elastostatics. Finite elasticity theory covers the simplest case where internal forces only depend on the current deformation of the material and not on its history, and is based on average data values. Within this framework, \textit{hyperelastic materials} are the class of material models described by a strain-energy function characterised by a set of \textit{deterministic} model parameters. In addition, for solid elastic materials, uncertainties in the observational data generally arise from the inherent variation in material properties and testing protocols \cite{Bayes:1763,Farmer:2017,Hughes:2010:HH,Oden:2018}. In view of these uncertainties, recently, stochastic representations of isotropic incompressible hyperelastic materials characterised by a stochastic strain-energy function, for which the model parameters are random variables following standard probability laws, were proposed in \cite{Staber:2015:SG}, while compressible versions of these models were constructed in \cite{Staber:2016:SG}. Ogden-type stochastic strain-energy functions were calibrated to experimental data for rubber and soft tissue materials in \cite{Mihai:2018a:MWG,Staber:2017:SG}, and anisotropic stochastic models with the model parameters as spatially-dependent random field variables were calibrated to vascular tissue data in \cite{Staber:2018:SG}. These models employ the maximum entropy principle for a discrete probability distribution introduced by Jaynes (1957) \cite{Jaynes:1957a,Jaynes:1957b,Jaynes:2003} and based on the notion of  entropy (or uncertainty) defined by Shannon (1948) \cite{Shannon:1948,Soni:2017:SG}. Such models can be useful for stochastic finite element implementations \cite{ArreguiMena:2016:AMMM,Babuska:2005:BTZ,Hauseux:2017:HHB,Hauseux:2018:HHCB}.

For stochastic hyperelastic models, the immediate question is: \emph{what is the influence of the random model parameters on the predicted nonlinear elastic responses?} This question was previously considered by us in \cite{Mihai:2018b:MWG}, for the stochastic Rivlin cube, and in \cite{Mihai:2018:MDWG}, for the symmetric inflation of internally pressurised stochastic spherical shells and tubes. These idealised problems illustrate some important effects on the likely elastic responses of stochastic hyperelastic materials under large strains.

Here, we address this question by employing a similar approach as in \cite{Mihai:2018b:MWG,Mihai:2018:MDWG} to revisit, in the context of stochastic elasticity, the cavitation problem of incompressible spheres of stochastic homogeneous isotropic hyperelastic materials under uniform radial tensile dead loads. Moreover, for all homogeneous isotropic hyperelastic models considered so far in the literature, cavitation appears as a supercritical bifurcation, where typically, after bifurcation, the cavity radius monotonically increases as the applied load increases (see, e.g., \cite{ChouWang:1989:CWH}). However, as we demonstrate here, the usual restriction that a material satisfies the Baker-Ericksen (BE) inequalities \cite{BakerEricksen:1954} is not sufficient to exclude the possibility of a subcritical bifurcation. In this case, one expects a \textit{snap cavitation} for which there is a jump in the radius of the cavity immediately after bifurcation. Indeed, we obtain the general conditions under which a cavitation can appear through a supercritical or subcritical bifurcation and construct explicitly, for the first time, examples of isotropic incompressible hyperelastic models that exhibit snap cavitation. The stochastic version of these models are then explored. In this case, we find that, due to the probabilistic nature of the model parameters, supercritical or subcritical bifurcation occurs with a given probability, and there is also a probability that the cavity may form under smaller or greater loads that the expected critical value. We refer to these phenomena as `likely cavitation'.

We begin, in Section~\ref{sec:models}, with a detailed presentation of the stochastic elastic framework. Then, in Section~\ref{sec:sphere}, for the stochastic sphere, after we review the elastic solution to the cavitation problem under uniformly applied tensile dead load, we recast the problem in the stochastic setting, and find the probabilistic solution. Concluding remarks are provided in Section~\ref{sec:conclude}.

%%%%%%%%%%%%%%%%%%%%%%%%%%%%%%%%%%%%%%%%%%%%%%%%%%%%%%%%%%%%
%%%%%%%%%%%%%%%%%%%%   NEW SECTION  %%%%%%%%%%%%%%%%%%%%%%%%
%%%%%%%%%%%%%%%%%%%%%%%%%%%%%%%%%%%%%%%%%%%%%%%%%%%%%%%%%%%%
\section{Stochastic isotropic hyperelastic models}\label{sec:models}

We recall that a homogeneous hyperelastic model is described by a strain-energy function $W(\mathbf{F})$ that depends on the deformation gradient tensor, $\mathbf{F}$, with respect to a fixed reference configuration, and is characterised by a set of \textit{deterministic} model parameters \cite{goriely17,Ogden:1997,TruesdellNoll:2004}. In contrast, a stochastic homogeneous hyperelastic model is defined by a stochastic strain-energy function, for which the model parameters are \emph{random variables} that satisfy standard probability laws \cite{Mihai:2018a:MWG,Staber:2015:SG,Staber:2016:SG,Staber:2017:SG}. In this case, each model parameter is described in terms of its \emph{mean value} and its \emph{variance}, which contains information about the range of values about the mean value. While it is rarely possible if ever to obtain complete information about a random quantity in an elastic sample of material, the partial information provided by the mean value and the variance is the most commonly used in many practical applications \cite{Caylak:2018:etal,Hughes:2010:HH,McCoy:1973}. Here, we combine finite elasticity  and information theory, and rely on the following general hypotheses \cite{Mihai:2018a:MWG,Mihai:2018b:MWG,Mihai:2018:MDWG}:
\begin{itemize}
\item[(A1)] Material objectivity: The principle of material objectivity (frame indifference) states that constitutive equations must be invariant under changes of frame of reference. It requires that the scalar strain-energy function, $W=W(\textbf{F})$, depending only on the deformation gradient $\textbf{F}$, with respect to the reference configuration, is unaffected by a superimposed rigid-body transformation (which involves a change of position) after deformation, i.e., $W(\textbf{R}^{T}\textbf{F})=W(\textbf{F})$, where $\textbf{R}\in SO(3)$ is a proper orthogonal tensor (rotation). Material objectivity is guaranteed by considering strain-energy functions defined in terms of invariants.
	
\item[(A2)] Material isotropy: The principle of isotropy requires that the strain-energy function is unaffected by a superimposed rigid-body transformation prior to deformation, i.e., $W(\textbf{F}\textbf{Q})=W(\textbf{F})$, where $\textbf{Q}\in SO(3)$. For isotropic materials, the  strain-energy  function is a symmetric function of the  principal stretches $\{\lambda_{i}\}_{i=1,2,3}$ of $\textbf{F}$, i.e., $W(\textbf{F})=\mathcal{W}(\lambda_{1},\lambda_{2},\lambda_{3})$.
	
\item[(A3)] Baker-Ericksen inequalities: In addition to the fundamental principles of objectivity and material symmetry, in order for the behaviour of a hyperelastic material to be physically realistic, there are some universally accepted constraints on the constitutive equations. Specifically, for a hyperelastic body, the Baker-Ericksen (BE) inequalities, which state that \emph{the greater principal (Cauchy) stress occurs in the direction of the greater principal stretch}, are \cite{BakerEricksen:1954}:
\begin{equation}\label{eq:BE}
\left(T_{i}-T_{j}\right)\left(\lambda_{i}-\lambda_{j}\right)>0\quad \mbox{if}\quad \lambda_{i}\neq\lambda_{j},\quad i,j=1,2,3,
\end{equation}
where $\{\lambda_{i}\}_{i=1,2,3}$ and $\{T_{i}\}_{i=1,2,3}$ denote the principal stretches and the principal Cauchy stresses, respectively, and ``$\geq$'' replaces the strict inequality ``$>$'' if any two principal stretches are equal. The BE inequalities \eqref{eq:BE} take the equivalent form \cite{BakerEricksen:1954,Marzano:1983}
\begin{equation}\label{eq:W:BE}
\left(\lambda_{i}\frac{\partial\mathcal{W}}{\partial\lambda_{i}}-\lambda_{j}\frac{\partial\mathcal{W}}{\partial\lambda_{j}}\right)\left(\lambda_{i}-\lambda_{j}\right)>0\quad \mbox{if}\quad \lambda_{i}\neq\lambda_{j},\quad i,j=1,2,3,
\end{equation}
where the strict inequality ``$>$'' is replaced by ``$\geq$'' if any two principal stretches are equal.

\item[(A4)] Finite mean and variance for the random shear modulus: We assume that, for any given finite deformation, the random shear modulus, $\mu$, and its inverse, $1/\mu$, are second-order random variables, i.e., they have finite mean value and finite variance \cite{Staber:2015:SG,Staber:2016:SG,Staber:2017:SG}.
\end{itemize}
While (A4) contains physically realistic expectations on the random shear modulus, which will be drawn from a probability distribution, assumptions (A1)-(A3) are well-known principles in isotropic finite elasticity \cite{goriely17,Ogden:1997,TruesdellNoll:2004}.

Specifically, we focus our attention on homogeneous incompressible hyperelastic materials characterised by the following stochastic strain-energy function \cite{Staber:2015:SG,Staber:2017:SG,Mihai:2018a:MWG},
\begin{equation}\label{eq:W:stoch}
	\mathcal{W}(\lambda_{1},\lambda_{2},\lambda_{3})=\frac{\mu_{1}}{2m^2}\left(\lambda_{1}^{2m}+\lambda_{2}^{2m}+\lambda_{3}^{2m}-3\right)
	+\frac{\mu_{2}}{2n^2}\left(\lambda_{1}^{2n}+\lambda_{2}^{2n}+\lambda_{3}^{2n}-3\right),
\end{equation}
where $m$ and $n$ are deterministic constants, and $\mu_{1}$ and $\mu_{2}$ are random variables following given probability distributions. In the deterministic elastic case,  $\mu_{1}$, $\mu_{2}$, $m$ and $n$ are constants, and the model contains, as special cases, the neo-Hookean model, the Mooney-Rivlin model, and the one- and two-term Ogden models. In both the deterministic elastic and stochastic cases, the shear modulus for infinitesimal deformations of these models is defined as $\mu=\mu_{1}+\mu_{2}$ \cite{Mihai:2017:MG,Mihai:2018a:MWG}. Note that we could easily extend our description to include $m$ and $n$ as stochastic variables as well. However, increasing the complexity in this way is not relevant for the present discussion. Including additional sources of randomness is an avenue of future research.

As it is well known, the deformation of an homogeneous isotropic hyperelastic material under uniaxial tension is a simple extension in the direction of the tensile force if and only if the BE inequalities hold \cite{Marzano:1983}. Under these conditions, the shear modulus is positive, but the individual coefficients may be either positive or negative, allowing for some interesting nonlinear elastic effects to be captured (see \cite{Mihai:2011:MG,Mihai:2013:MG,Mihai:2017:MG,Mihai:2018b:MWG} and the references therein). In particular, in the present paper, the initiation of either stable or unstable snap cavitation in a homogeneous isotropic sphere will be presented.

For the stochastic materials described by \eqref{eq:W:stoch}, condition (A4) is guaranteed by setting the following mathematical expectations \cite{Staber:2015:SG,Staber:2016:SG,Staber:2017:SG,Mihai:2018a:MWG,Mihai:2018b:MWG}:
\begin{eqnarray}\label{eq:Emu1}\begin{cases}
		E\left[\mu\right]=\underline{\mu}>0,&\\
		E\left[\log\ \mu\right]=\nu,& \mbox{such that $|\nu|<+\infty$}.\label{eq:Emu2}\end{cases}
\end{eqnarray}
Then, under the constraints (\ref{eq:Emu1}), the random shear modulus, $\mu$, with mean value $\underline{\mu}$ and standard deviation $\|\mu\|=\sqrt{\text{Var}[\mu]}$, defined as the square root of the variance, $\text{Var}[\mu]$, follows a Gamma probability distribution \cite{Soize:2000,Soize:2001}, with hyperparameters $\rho_{1}>0$ and $\rho_{2}>0$ satisfying
\begin{equation}\label{eq:rho12}
	\underline{\mu}=\rho_{1}\rho_{2},\qquad
	\|\mu\|=\sqrt{\rho_{1}}\rho_{2}.
\end{equation}
The corresponding probability density function takes the form \cite{Abramowitz:1964,Johnson:1994:JKB}
\begin{equation}\label{eq:mu:gamma}
	g(\mu;\rho_{1},\rho_{2})=\frac{\mu^{\rho_{1}-1}e^{-\mu/\rho_{2}}}{\rho_{2}^{\rho_{1}}\Gamma(\rho_{1})},\qquad\mbox{for}\ \mu>0\ \mbox{and}\ \rho_{1}, \rho_{2}>0,
\end{equation}
where $\Gamma:\mathbb{R}^{*}_{+}\to\mathbb{R}$ is the complete Gamma function
\begin{equation}\label{eq:gamma}
	\Gamma(z)=\int_{0}^{+\infty}t^{z-1}e^{-t}\text dt.
\end{equation}
For technical convenience, we set a finite constant value $b>-\infty$, such that $\mu_{i}>b$, $i=1,2$ (e.g., $b=0$ if $\mu_{1}>0$ and $\mu_{2}>0$, but $b$ is not unique in general), and introduce the auxiliary random variable \cite{Mihai:2018a:MWG}
\begin{equation}\label{eq:R12:b}
R_{1}=\frac{\mu_{1}-b}{\mu-2b},
\end{equation}
such that $0<R_{1}<1$. Consequently, we can equivalently express the random model parameters $\mu_{1}$ and $\mu_{2}$ as follows,
\begin{equation}\label{eq:mu12:b}
\mu_{1}=R_{1}(\mu-2b)+b,\qquad \mu_{2}=\mu-\mu_{1}=(1-R_{1})(\mu-2b)+b.
\end{equation}
It is reasonable to assume \cite{Staber:2015:SG,Staber:2016:SG,Staber:2017:SG,Mihai:2018a:MWG}
\begin{eqnarray}\begin{cases}
E\left[\log\ R_{1}\right]=\nu_{1},& \mbox{such that $|\nu_{1}|<+\infty$},\label{eq:ER1}\\
E\left[\log(1-R_{1})\right]=\nu_{2},& \mbox{such that $|\nu_{2}|<+\infty$},\label{eq:ER2}\end{cases}
\end{eqnarray}
in which case, the random variable $R_{1}$, with mean value $\underline{R}_{1}$ and variance $\text{Var}[R_{1}]$, follows a standard Beta distribution \cite{Abramowitz:1964,Johnson:1994:JKB}, with hyperparameters $\xi_{1}>0$ and $\xi_{2}>0$ satisfying
\begin{equation}\label{eq:xi12}
\underline{R}_{1}=\frac{\xi_{1}}{\xi_{1}+\xi_{2}},\qquad
\text{Var}[R_{1}]=\frac{\xi_{1}\xi_{2}}{\left(\xi_{1}+\xi_{2}\right)^2\left(\xi_{1}+\xi_{2}+1\right)}.
\end{equation}
The associated probability density function is
\begin{equation}\label{eq:betaR1}
\beta(r;\xi_{1},\xi_{2})=\frac{r^{\xi_{1}-1}(1-r)^{\xi_{2}-1}}{B(\xi_{1},\xi_{2})},\qquad \qquad\mbox{for}\ r\in(0,1)\ \mbox{and}\ \xi_{1}, \xi_{2}>0,
\end{equation}
where $B:\mathbb{R}^{*}_{+}\times\mathbb{R}^{*}_{+}\to\mathbb{R}$ is the Beta function
\begin{equation}\label{eq:beta}
B(x,y)=\int_{0}^{1}t^{x-1}(1-t)^{y-1}dt.
\end{equation}
Thus, for the random coefficients given by \eqref{eq:mu12:b}, the corresponding mean values take the form,
\begin{equation}\label{eq:mu12:mean}
\underline{\mu}_{1}=\underline{R}_{1}(\underline{\mu}-2b)+b,
\qquad \underline{\mu}_{2}=\underline{\mu}-\underline{\mu}_{1}=(1-\underline{R}_{1})(\underline{\mu}-2b)+b,
\end{equation}
and the variances and covariance are, respectively,
\begin{eqnarray}
&&\text{Var}\left[\mu_{1}\right]
=(\underline{\mu}-2b)^2\text{Var}[R_{1}]+(\underline{R}_{1})^2\text{Var}[\mu]+\text{Var}[\mu]\text{Var}[R_{1}],\\
&&\text{Var}\left[\mu_{2}\right]
=(\underline{\mu}-2b)^2\text{Var}[R_{1}]+(1-\underline{R}_{1})^2\text{Var}[\mu]+\text{Var}[\mu]\text{Var}[R_{1}],\\
&&\text{Cov}[\mu_{1},\mu_{2}]=\frac{1}{2}\left(\text{Var}[\mu]-\text{Var}[\mu_{1}]-\text{Var}[\mu_{2}]\right).
\end{eqnarray}
It should be noted that the random variables $\mu$ and $R_1$ are independent, depending on parameters $(\rho_1,\rho_2)$ and $(\zeta_1,\zeta_2)$, respectively, which are derived by fitting distributions to given data. However, $\mu_1$ and $\mu_2$ are dependent variables as they both require $(\mu,R_1)$ to be defined. Explicit derivations of the probability distributions for the random parameters when stochastic isotropic hyperelastic models are calibrated to experimental data are presented in \cite{Mihai:2018a:MWG,Staber:2017:SG}.

Our aim here is to analyse the radially symmetric finite deformations of a sphere of stochastic hyperelastic material defined by \eqref{eq:W:stoch}, under tension, when subject to prescribed surface dead loads applied uniformly in the radial direction. One can view the stochastic sphere as an ensemble (or population) of  spheres, where each sphere has the same initial radius and is made from a homogeneous isotropic incompressible hyperelastic material, with the elastic parameters not known with certainty, but drawn from known probability distributions. Then, for every hyperelastic sphere, the finite elasticity theory applies. For the stochastic hyperelastic body, the question is: \emph{what is the probability distribution of stable radially symmetric deformation under a given surface dead load?}

%%%%%%%%%%%%%%%%%%%%%%%%%%%%%%%%%%%%%%%%%%%%%%%%%%%%%%%%%%%%
%%%%%%%%%%%%%%%%%%%%   NEW SECTION  %%%%%%%%%%%%%%%%%%%%%%%%
%%%%%%%%%%%%%%%%%%%%%%%%%%%%%%%%%%%%%%%%%%%%%%%%%%%%%%%%%%%%
\section{Incompressible spheres}\label{sec:sphere}

In this section, we consider a sphere of stochastic incompressible hyperelastic material described by \eqref{eq:W:stoch}, subject to a radially symmetric deformation, caused by the sole action of a given radial tensile dead load. As for the deterministic elastic sphere \cite{Ball:1982}, we obtain conditions on the constitutive law, such that, setting the internal pressure equal to zero, where the radius tends to zero, the required external dead load is finite, and therefore cavitation occurs. We further analyse the stability of the cavitated solution, and distinguish between supercritical cavitation, where the cavity radius monotonically increases as the dead load increases, and subcritical (snap) cavitation, with a sudden jump to a finite internal radius immediately after initiation. To the best of our knowledge, in the deterministic elastic case, the onset of snap cavitation  in a homogeneous isotropic sphere has not been discussed before. Therefore, we start our analysis in the deterministic elastic context before  extending it to the stochastic case.

\begin{figure}[htbp]
	\begin{center}
		\includegraphics[width=0.65\textwidth]{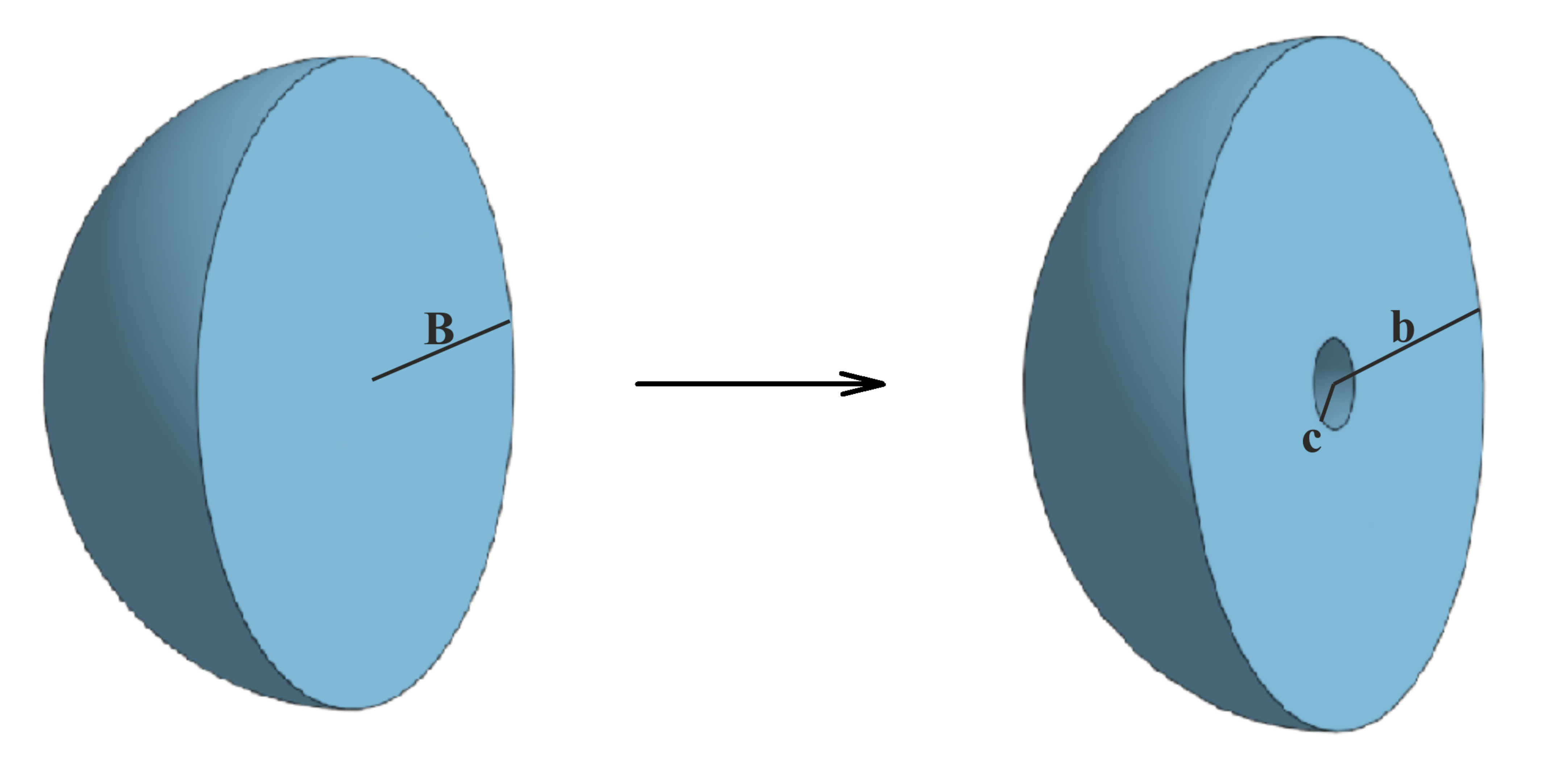}
		\caption{Schematic of cross-section of a sphere, showing the reference state, with outer radius $B$ (left), and the deformed state, with cavity radius $c$ and outer radius $b$ (right), respectively.}\label{fig:hsphere}
	\end{center}
\end{figure}

For the stochastic sphere, the radially symmetric deformation takes the form
\begin{equation}\label{eq:sphere:deform}
r=g(R),\qquad \theta=\Theta,\qquad \phi=\Phi,
\end{equation}
where $(R,\Theta,\Phi)$ and $(r,\theta,\phi)$ are the spherical polar coordinates in the reference and current configuration, respectively, such that $0\leq R\leq B$, and $g(R)\geq 0$ is to be determined. The corresponding deformation gradient is equal to $\textbf{F}=\text{diag}\left(\lambda_{1},\lambda_{2},\lambda_{3}\right)$, with
\begin{equation}\label{eq:sphere:lambdas:g}
\lambda_{1}=\frac{\text{d}g}{\text{d}R}=\lambda^{-2},\qquad \lambda_{2}=\lambda_{3}=\frac{g(R)}{R}=\lambda,
\end{equation}
where $\lambda_{1}$ and  $\lambda_{2}=\lambda_{3}$ are the radial and hoop stretches, respectively, and $\text{d}g/\text{d}R$ denotes the derivative of $g$ with respect to $R$. By \eqref{eq:sphere:lambdas:g},
\begin{equation}\label{eq:sphere:dg}
g^2\frac{\text{d}g}{\text{d}R}=R^2,
\end{equation}
hence,
\begin{equation}\label{eq:sphere:g}
g(R)=\left(R^3+c^3\right)^{1/3},
\end{equation}
where $c\geq 0$ is a constant to be calculated. If $c>0$, then $g(R)\to c>0$ as $R\to 0_{+}$, and a spherical cavity of radius $c$ forms at the centre of the sphere, from zero initial radius (see Figure~\ref{fig:hsphere}), otherwise the sphere remains undeformed.

%%%%%%%%%%%%%%%
%%%%%%%%%%%%%%%

Assuming that the deformation \eqref{eq:sphere:deform} is due to a prescribed radial tensile dead load, applied uniformly on the sphere surface in the reference configuration, in the absence of body forces, the radial equation of equilibrium is
\begin{equation}\label{eq:equilib}
\frac{dP_{11}}{\text{d}R}+\frac{2}{R}(P_{11}-P_{22})=0,
\end{equation}
or equivalently,
\begin{equation}\label{eq:sphere:equilib}
\frac{dP_{11}}{\text{d}\lambda}\lambda^{-2}+2\frac{P_{11}-P_{22}}{1-\lambda^{3}}=0,
\end{equation}
where $\textbf{P}=(P_{ij})_{i,j=1,2,3}$ is the first Piola-Kirchhoff stress tensor. For an incompressible material,
\[
P_{11}=\frac{\partial\mathcal{W}}{\partial\lambda_{1}}-\frac{p}{\lambda_{1}},\qquad
P_{22}=\frac{\partial\mathcal{W}}{\partial\lambda_{2}}-\frac{p}{\lambda_{2}}.
\]
Denoting
\begin{equation}\label{eq:sphere:W}
W(\lambda)=\mathcal{W}(\lambda^{-2},\lambda,\lambda),
\end{equation}
where $\lambda=r/R=g(R)/R=(1+c^3/R^3)^{1/3}>1$, we obtain
\begin{equation}\label{eq:sphere:dW}
\frac{\text{d}W}{\text{d}\lambda}=-\frac{2}{\lambda^3}\frac{\partial\mathcal{W}}{\partial\lambda_{1}}+2\frac{\partial\mathcal{W}}{\partial\lambda_{2}}
=-\frac{2P_{11}}{\lambda^3}+2P_{22}.
\end{equation}
Then, setting the internal pressure (at $R\to 0_{+}$) equal to zero, by \eqref{eq:sphere:equilib} and \eqref{eq:sphere:dW}, the external tension (at $R=B$) is equal to
\begin{equation}\label{eq:sphere:integral:T}
T=\frac{P_{11}}{\lambda^2}\left|_{\lambda=\lambda_{b}}\right.=\int_{\lambda_{b}}^{\lambda_{c}}\frac{\text{d}W}{\text{d}\lambda}\frac{\text{d}\lambda}{\lambda^3-1},
\end{equation}
and the applied dead load, in the reference configuration, is
\begin{equation}\label{eq:sphere:integral:P}
P=T\lambda_{b}^2=\lambda_{b}^2\int_{\lambda_{b}}^{\lambda_{c}}\frac{\text{d}W}{\text{d}\lambda}\frac{\text{d}\lambda}{\lambda^3-1},
\end{equation}
where $\lambda_{c}$ and $\lambda_{b}$ represent the stretches at the centre and outer surface, respectively. The value of the required dead load, $P_{0}$, for the onset of cavitation (bifurcation from the reference state) is obtained by taking $\lambda_{c}\to\infty$ and $\lambda_{b}=\left(1+c^3/B^3\right)^{1/3}\to 1$ as $c\to 0_{+}$ in \eqref{eq:sphere:integral:P}, i.e.,
\begin{equation}\label{eq:sphere:integral:P0}
P_{0}=\int_{1}^{\infty}\frac{\text{d}W}{\text{d}\lambda}\frac{\text{d}\lambda}{\lambda^3-1}.
\end{equation}
The BE inequalities \eqref{eq:W:BE} imply
\begin{equation}\label{eq:sphere:W:BE}
\frac{\text{d}W}{\text{d}\lambda}\frac{1}{\lambda^3-1}>0,
\end{equation}
hence, $P_{0}>0$. Then, if the critical dead load given by \eqref{eq:sphere:integral:P0} is finite, cavitation takes place, else, the sphere remains undeformed.

Before considering the stochastic setting, we briefly revisit the deterministic elastic case.

%%%%%%%%%%%%%%%%%%%%   SUB-SECTION  %%%%%%%%%%%%%%%%%%%%%%%%
\subsection{Deterministic elastic spheres}

For a sphere made of a hyperelastic material with the strain-energy function
\begin{equation}\label{eq:W}
\mathcal{W}(\lambda_{1},\lambda_{2},\lambda_{3})=\frac{\mu_{1}}{2m^2}\left(\lambda_{1}^{2m}+\lambda_{2}^{2m}+\lambda_{3}^{2m}-3\right)
+\frac{\mu_{2}}{2n^2}\left(\lambda_{1}^{2n}+\lambda_{2}^{2n}+\lambda_{3}^{2n}-3\right),
\end{equation}
where $\mu_{1}$ and $\mu_{2}$ are positive constants,  \eqref{eq:sphere:W} takes the form
\begin{equation}\label{eq:sphere:W:0}
W(\lambda)=\frac{\mu_{1}}{2m^2}\left(\lambda^{-4m}+2\lambda^{2m}-3\right)+\frac{\mu_{2}}{2n^2}\left(\lambda^{-4n}+2\lambda^{2n}-3\right).
\end{equation}
For the onset of cavitation, the critical dead load traction, defined by \eqref{eq:sphere:integral:P0}, is equal to
\begin{equation}\label{eq:sphere:integral:P0:fe}
P_{0}=\frac{2\mu_{1}}{m}\int_{1}^{\infty}\frac{\lambda^{2m-1}-\lambda^{-4m-1}}{\lambda^3-1}d\lambda+\frac{2\mu_{2}}{n}\int_{1}^{\infty}\frac{\lambda^{2n-1}-\lambda^{-4n-1}}{\lambda^3-1}d\lambda,
\end{equation}
or equivalently, by the change of variable $x=\lambda^{3}-1$,
\begin{equation}\label{eq:sphere:integral:P0:x}
\begin{split}
P_{0}&=\frac{2\mu_{1}}{3m}\int_{0}^{\infty}\frac{(x+1)^{(2m-3)/3}-(x+1)^{-(4m+3)/3}}{x}dx\\
	&+\frac{2\mu_{2}}{3n}\int_{0}^{\infty}\frac{(x+1)^{(2n-3)/3}-(x+1)^{-(4n+3)/3}}{x}dx.
\end{split}
\end{equation}
By \eqref{eq:sphere:integral:P0:x}, $P_{0}$ is finite, and hence, a spherical cavity forms, if and only if the following conditions are simultaneously satisfied:  $2m-3<0$, $-4m-3<0$, $2n-3<0$, $-4n-3<0$, or equivalently  \cite{ChouWang:1989:CWH,Horgan:1995:HP} (see also Example 5.1 of \cite{Ball:1982}), if and only if
\begin{equation}\label{eq:sphere:mnbounds}
-3/4< m, n<3/2.
\end{equation}
In particular, cavitation is found in a neo-Hookean sphere (with $m=1$ and $n=0$), but not in a Mooney-Rivlin sphere (with $m=1$ and $n=-1$). The special cases when $m\in\{-1/2, 1\}$ and $n=0$, are given as examples in \cite{Ball:1982}, and when $m\in\{1/2, 3/4, 1, 5/4\}$ and $n=0$, the explicit critical loads are provided in \cite{ChouWang:1989:CWH}. When these bounds and the BE inequalities are satisfied, the critical pressure $P_{0}$ is finite and the problem is to find the behavior of the cavity in a neighborhood of this critical value. In each of those previously studied cases (see, e.g., Figure~2 of \cite{ChouWang:1989:CWH}), cavitation forms from zero radius and then presents itself  as a supercritical bifurcation with stable cavitation (i.e. the new bifurcated solution exists locally for values of $P>P_{0}$, and the radius of the cavity monotonically increases with the applied load post-bifurcation).

Another theoretical possibility is that the bifurcation could be subcritical (i.e., the cavitated solution exists locally for values less than $P_{0}$ and is unstable). One would then expect an unstable \textit{snap cavitation} with a sudden jump to a cavitated solution with a finite internal radius. This subcritical behaviour of the homogeneous isotropic elastic sphere has not been explicitly demonstrated in the literature before. Here, we show that, depending on the model parameters, the family of materials~\eqref{eq:sphere:W} can exhibit both behaviours. General conditions for a given material to exhibit either a subcritical or supercritical bifurcation are provided in Appendix~\ref{sec:append}. 

%%%%%%%%%%%%%%%
\begin{figure}[htbp]
	\begin{center}
		\includegraphics[width=1\textwidth]{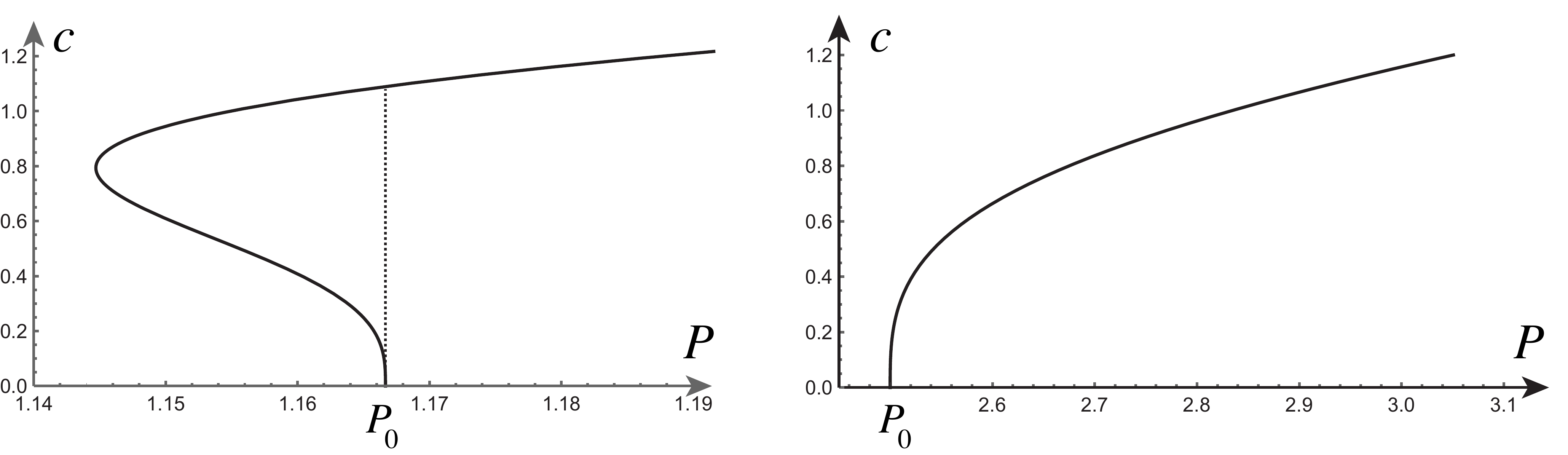}
		\caption{Subcritical (left) and supercitical (right) cavitation found in a unit sphere (with $B=1$) of material model~\eqref{eq:sphere:W:0} with $\mu_{1}=1$ and either $\mu=2/3$ (left) or $\mu=1$ (right). The dashed line indicates the snap cavitation expected at the bifurcation, leading to a sudden increase of the cavity size in the subcritical case.}\label{fig:subcav}
	\end{center}
	%\end{figure}
	%%%%%%%%%%%%%%%
	%\begin{figure}[htbp]
	\begin{center}
		\includegraphics[width=0.8\textwidth]{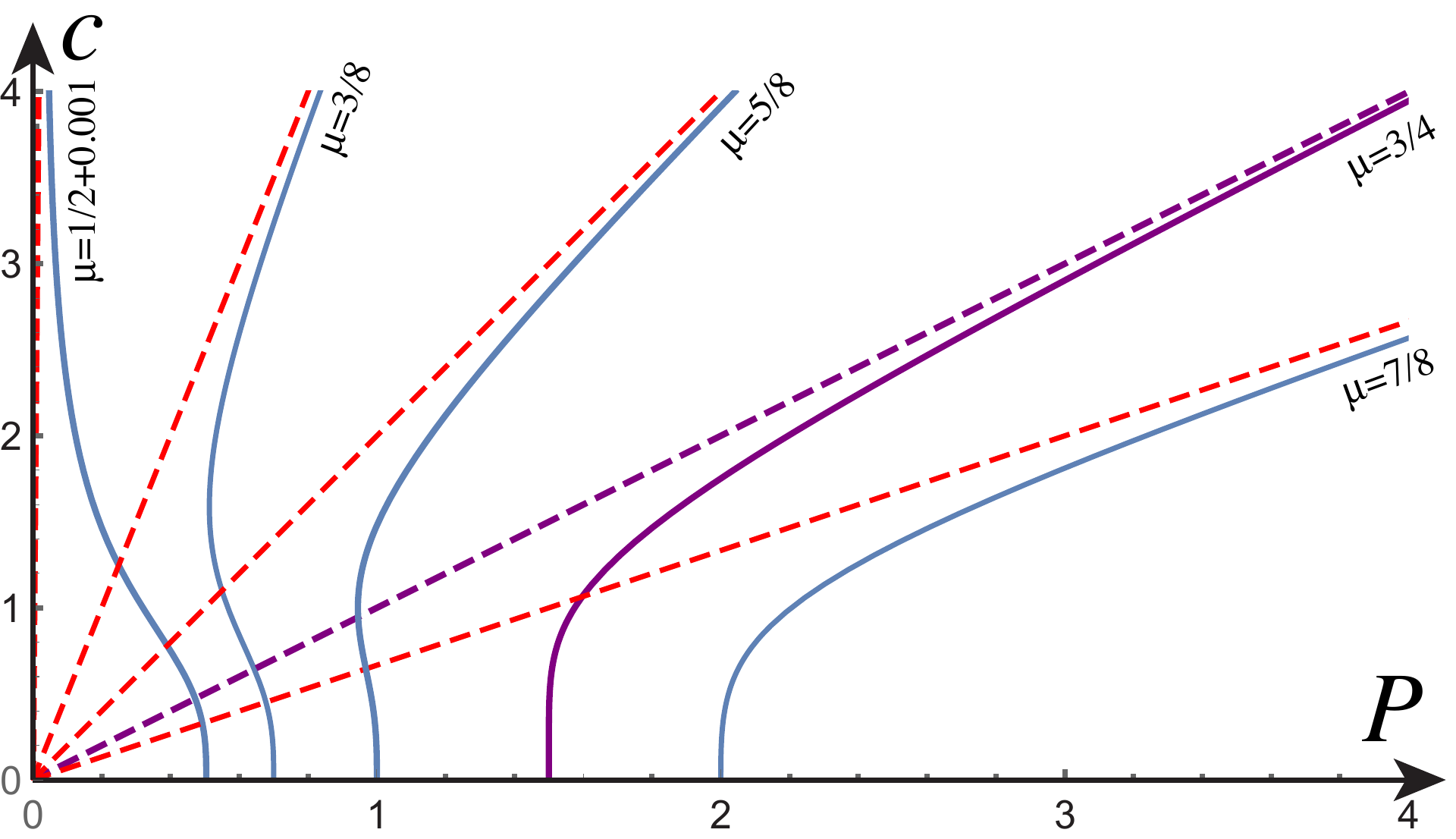}
		\caption{Change of behavior under various parameter values found in a unit sphere (with $B=1$) of material model~\eqref{eq:sphere:W:0} with $\mu_{1}=1$. Note the critical case at $\mu=3/4$. The dashed line indicates the asymptotic behavior for large values of $P$ and is given by $P=(4\mu-2\mu_{1})c/B$.}\label{fig:bif}
	\end{center}
\end{figure}
%%%%%%%%%%%%%%%%

As an example, we illustrate the variety of behaviours when $m=1$ and $n=-1/2$, such that \eqref{eq:sphere:W} takes the form
\begin{equation}\label{eq:sphere:W:ex}
W(\lambda)=\frac{\mu_{1}}{2}\left(\lambda^{-4}+2\lambda^{2}-3\right)+2\mu_{2}\left(\lambda^{2}+2\lambda^{-1}-3\right).
\end{equation}
In this case, under the deformation \eqref{eq:sphere:deform}, the BE inequalities \eqref{eq:W:BE} are reduced to
\begin{equation}\label{eq:sphere:BE}
\mu_{1}+2\mu_{2}\frac{\lambda^3}{1+\lambda^3}>0.
\end{equation}
The inequality \eqref{eq:sphere:BE} implies that, when $\lambda\to 1$, the  shear modulus must be positive, i.e., $\mu=\mu_{1}+\mu_{2}>0$, while if $\lambda\to\infty$, then $\mu_{1}+2\mu_{2}>0$. Noting that the function of $\lambda$ on the left-hand side is monotonically increasing when $\mu_{2}$ is positive, and decreasing if $\mu_{2}$ is negative, and taking $\mu_{1}>0$, the two limits imply that the BE inequalities are satisfied for all values of $\lambda$ if
\begin{equation}\label{eq:sphere:mu1mu:BE}
0<\frac{\mu_1}{\mu}<2.
\end{equation}
For sufficiently small $c/B$, the corresponding dead-load traction, defined by \eqref{eq:sphere:integral:P}, is equal to
\begin{equation}\label{eq:sphere:P:st}
\begin{split}
P&=2\mu_{1}\left[\left(1+\frac{c^3}{B^3}\right)^{1/3}+\frac{1}{4}\left(1+\frac{c^3}{B^3}\right)^{-2/3}\right]+4\mu_{2}\left(1+\frac{c^3}{B^3}\right)^{1/3}\\
&=4\mu\left(1+\frac{c^3}{B^3}\right)^{1/3}-2\mu_{1}\left[\left(1+\frac{c^3}{B^3}\right)^{1/3}-\frac{1}{4}\left(1+\frac{c^3}{B^3}\right)^{-2/3}\right].
\end{split}	
\end{equation}
Then, the critical tensile dead load given by \eqref{eq:sphere:integral:P0} takes the form
\begin{equation}\label{eq:sphere:P0:st}
P_{0}=4\mu-\frac{3\mu_{1}}{2}.
\end{equation}
As $P_{0}$ is positive in \eqref{eq:sphere:P0:st}, we have $0<\mu_{1}/\mu<8/3$, which is guaranteed by \eqref{eq:sphere:mu1mu:BE}.

The question is now to find the possible behaviour of the cavity opening $c$ as a function of $P$ in a neighbourhood of $P_{0}$. On differentiating \eqref{eq:sphere:P:st} with respect to $c/B$, we obtain
\begin{equation}\label{eq:sphere:dP:ex}
\frac{\text{d}P}{\text{d}(c/B)}=2\frac{c^2}{B^2}\left\{2\mu\left(1+\frac{c^3}{B^3}\right)^{-2/3}-\mu_{1}\left[\left(1+\frac{c^3}{B^3}\right)^{-2/3}+\frac{1}{2}\left(1+\frac{c^3}{B^3}\right)^{-5/3}\right]\right\}.
\end{equation}
Hence, by Proposition~\ref{app:prop} given in Appendix~\ref{sec:append} (with $n=3$), when
\begin{equation}\label{eq:sphere:mu1mu:stable}
0<\frac{\mu_{1}}{\mu}<\frac{4}{3}=\text{inf}_{0<c/B<1}\left[2\left(1+\frac{c^3}{B^3}\right)\left(\frac{3}{2}+\frac{c^3}{B^3}\right)^{-1}\right],
\end{equation}
where ``$\text{inf}$'' denotes infimum, the bifurcation is supercritical and the radius of the cavity monotonically increases as the tensile dead load increases. However, if there exists $c_{0}>0$, such that
\begin{equation}\label{eq:sphere:mu1mu:unstable}
2\left(1+\frac{c_{0}^3}{B^3}\right)\left(\frac{3}{2}+\frac{c_{0}^3}{B^3}\right)^{-1}<\frac{\mu_{1}}{\mu}<2,
\end{equation}
then the bifurcation is subcritical and the required applied load starts to decrease at $c=c_{0}$, where there is a sudden jump in the opening of cavity. In particular, if \eqref{eq:sphere:mu1mu:unstable} holds for $c_{0}=0$, i.e.,
\begin{equation}\label{eq:sphere:mu1mu:bounds}
\frac{4}{3}<\frac{\mu_{1}}{\mu}<2,
\end{equation}
then \eqref{eq:sphere:mu1mu:BE} is valid while the cavitation becomes unstable.

Thus, $dP/d(c/B)\to 0$ as $c\to 0_{+}$, and by Proposition~\ref{app:prop}, the bifurcation at the critical load, $P_{0}$, is supercritical (respectively, subcritical) if $dP/d(c/B)>0$ (respectively, $dP/d(c/B)<0$) for arbitrarily small $c/B$. Examples of both these behaviours are illustrated in Figures~\ref{fig:subcav} and~\ref{fig:bif}.

%%%%%%%%%%%%%%%%%%%%   SUB-SECTION  %%%%%%%%%%%%%%%%%%%%%%%%
\subsection{Stochastic elastic spheres}

We now turn our attention to the stochastic model described by \eqref{eq:W:stoch}, with $m=1$ and $n=-1/2$, and the other parameters drawn from probability distributions. In this case, recalling that $\mu$ follows a Gamma distribution $g(u;\rho_{1},\rho_{2})$, defined by \eqref{eq:mu:gamma}, the probability distribution of stable cavitation is equal to
\begin{equation}\label{eq:sphere:P1}
P_{1}(\mu_{1})=1-\int_{0}^{3\mu_{1}/4}g(u;\rho_{1},\rho_{2})du,
\end{equation}
and that of unstable cavitation is
\begin{equation}\label{eq:sphere:P2}
P_{2}(\mu_{1})=1-P_{1}(\mu_{1}).
\end{equation}

%%%%%%%%%%%%%%
\begin{figure}[h!!!tb]
	\begin{center}
		\includegraphics[width=0.5\textwidth]{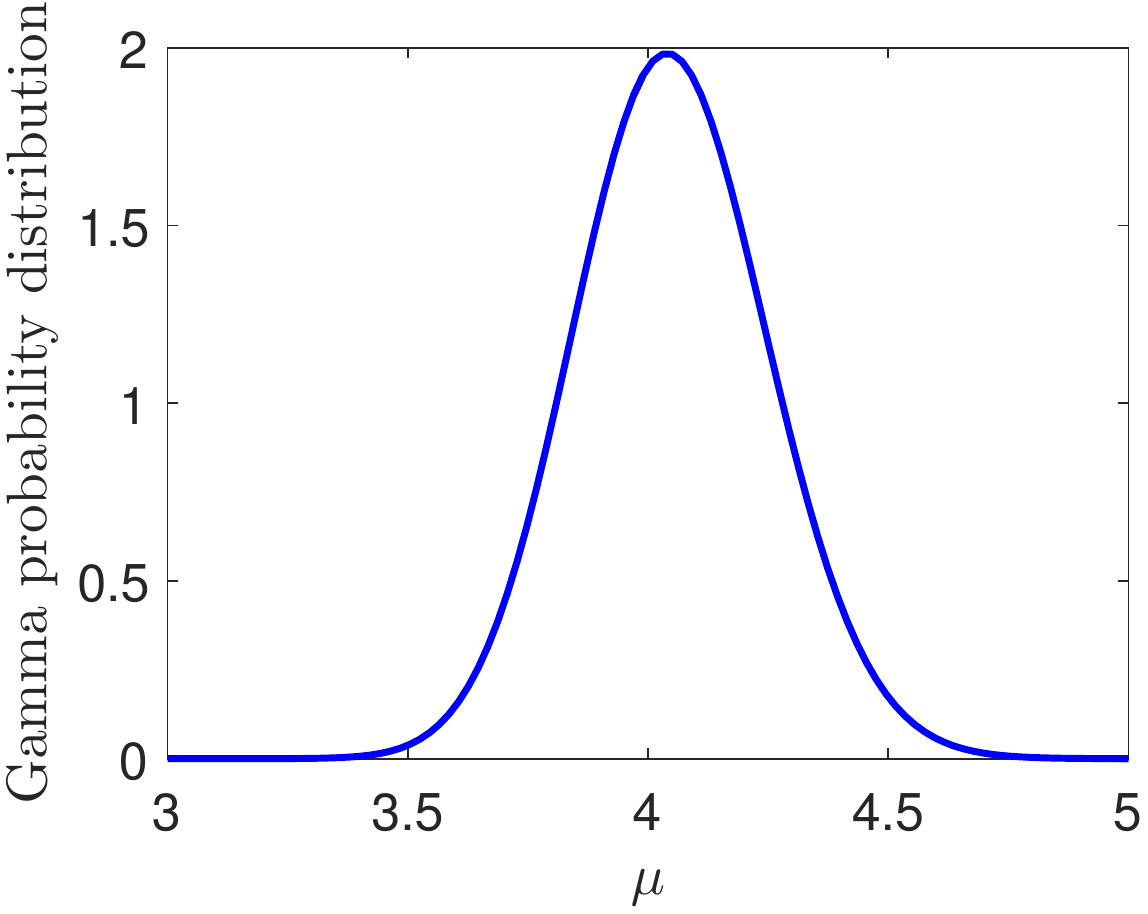}
		\caption{Example of Gamma distribution, with $\rho_1=405$ and $\rho_2=0.01$, for the random shear modulus $\mu>0$.}\label{fig:mu-gpdf}
	\end{center}
\end{figure}
%%%%%%%%%%%%%%

%%%%%%%%%%%%%%
\begin{figure}[h!!!tb]
	\begin{center}
		\includegraphics[width=\textwidth]{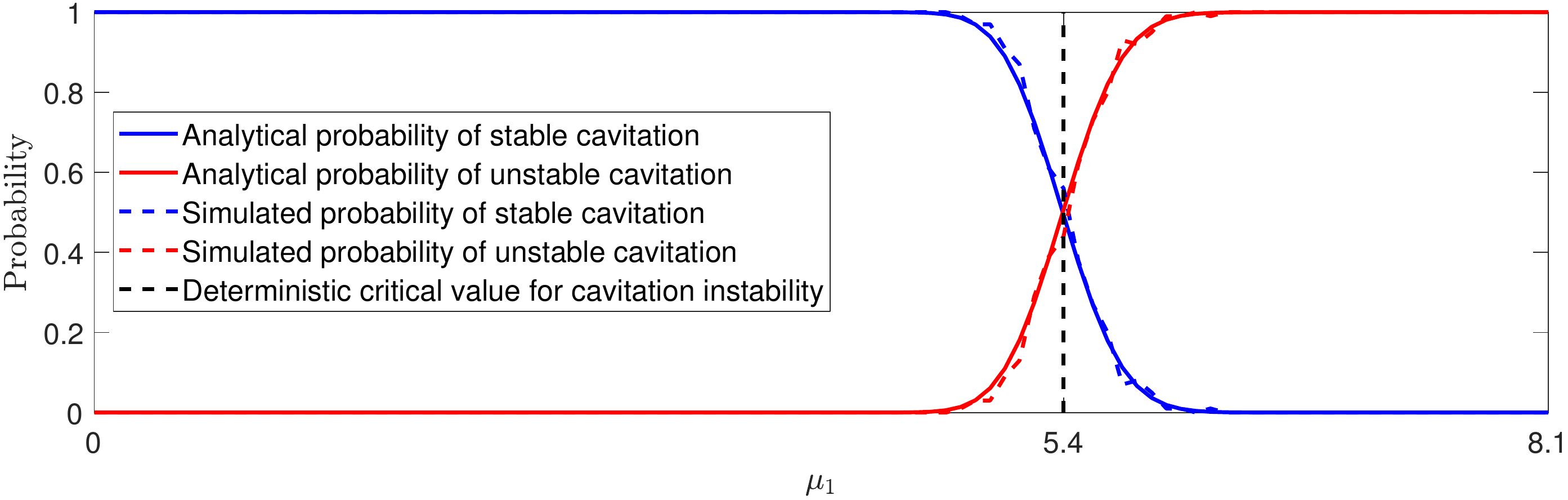}
		\caption{Probability distributions of whether cavitation is stable or not in a sphere of stochastic material described by \eqref{eq:W:stoch} with $m=1$ and $n=-1/2$, when the shear modulus, $\mu$, follows a Gamma distribution with $\rho_{1}=405$ and $\rho_{2}=0.01$. Continuous coloured lines represent analytically derived solutions, given by equations \eqref{eq:sphere:P1}-\eqref{eq:sphere:P2}, and the dashed versions represent stochastically generated data. The vertical line at the critical value, $4\underline{\mu}/3=5.4$, separates the expected regions based only on the mean value of the shear modulus, $\underline{\mu}=\rho_{1}\rho_{2}=4.05$.}\label{fig:sphere-pdfs}
	\end{center}
\end{figure}
%%%%%%%%%%%%%%%

%%%%%%%%%%%%%%
\begin{figure}[htbp]
	\begin{center}
		(a)\includegraphics[width=0.46\textwidth]{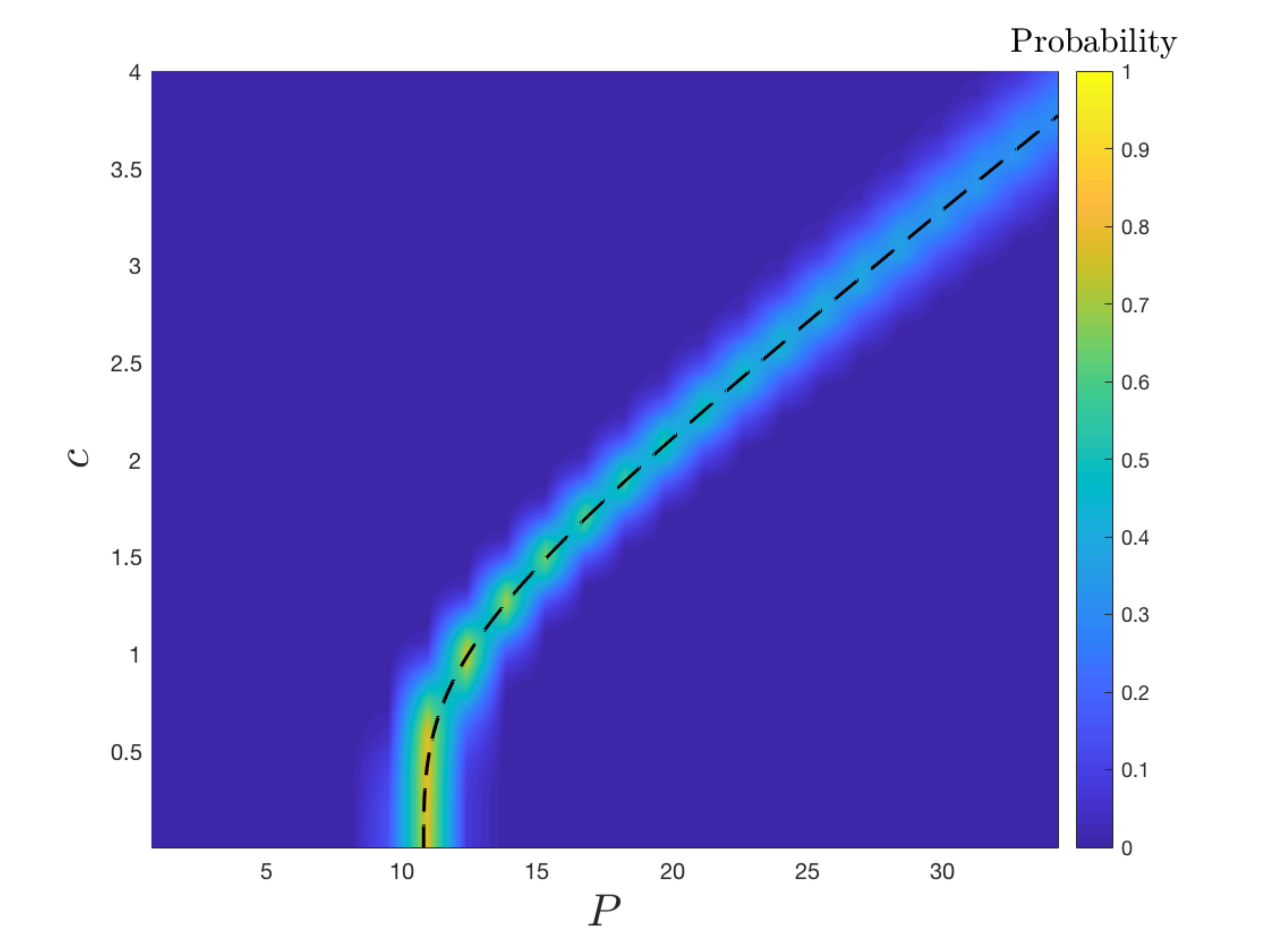}
		(b)\includegraphics[width=0.46\textwidth]{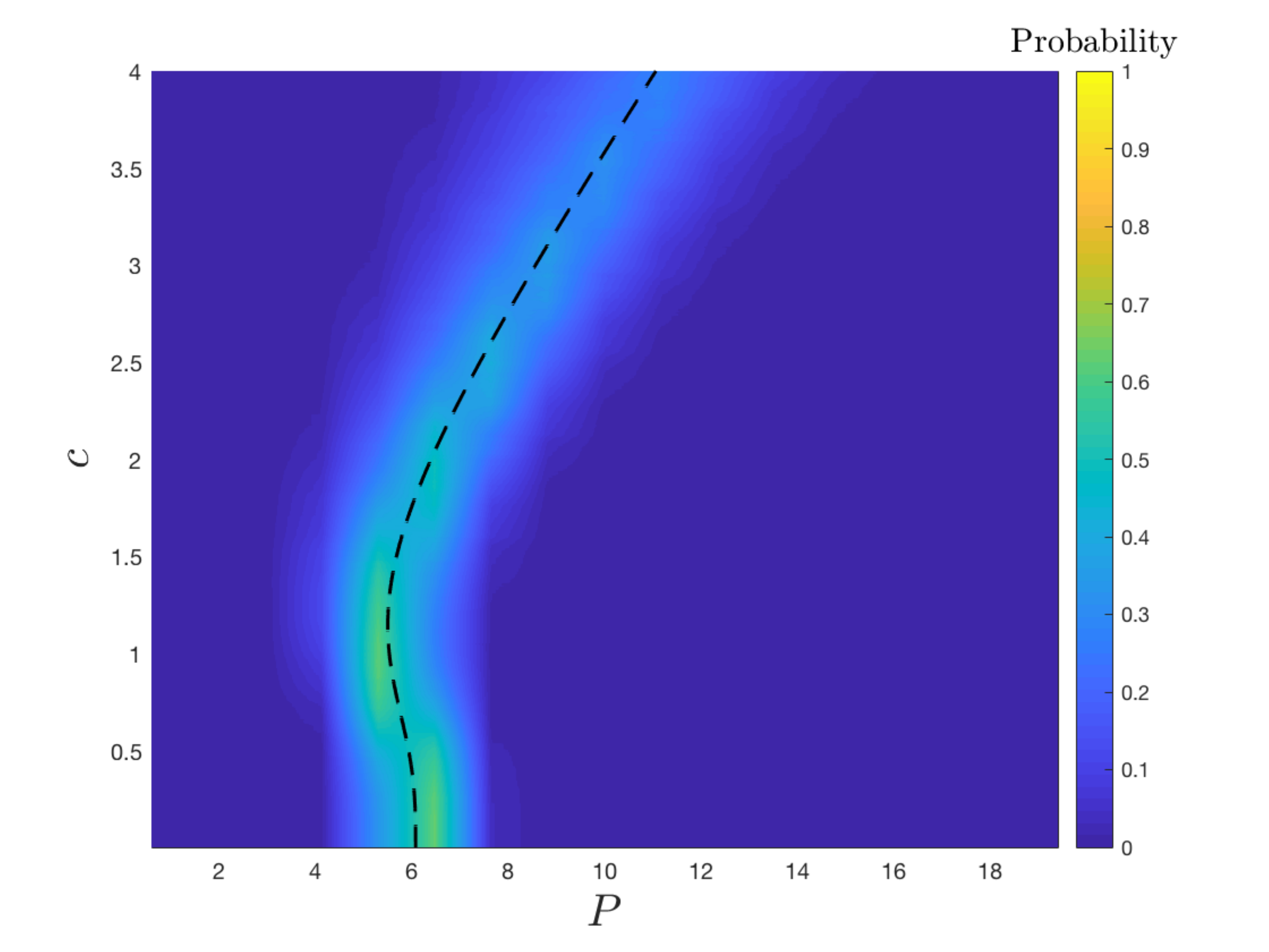}
		\caption{Probability distribution of the applied dead-load traction $P$ causing cavitation of radius $c$ in a unit sphere (where $B=1$) of stochastic material described by \eqref{eq:W:stoch} with $m=1$ and $n=-1/2$, when $\mu$ follows a Gamma distribution with $\rho_{1}=405$ and $\rho_{2}=0.01$, and: (a) $R_1=\mu_{1}/\mu$ follows a Beta distribution with $\xi_{1}=287$ and $\xi_{2}=36$; (b) $R_1=(\mu_1+3)/(\mu+6)$ follows a Beta distribution with $\xi_{1}=325$ and $\xi_{2}=10$. The dashed black lines correspond to the expected bifurcation based only on mean parameter values.}\label{fig:sphere-stdiag}
	\end{center}
\end{figure}
%%%%%%%%%%%%%%%

For example, taking $\rho_{1}=405$ and  $\rho_{2}= 0.01$ (see Figure~\ref{fig:mu-gpdf}), the mean value of the shear modulus is $\underline{\mu}=\rho_{1}\rho_{2}=4.05$, and the probability distributions given by equations \eqref{eq:sphere:P1}-\eqref{eq:sphere:P2} are illustrated numerically in Figure~\ref{fig:sphere-pdfs} (with blue lines for $P_{1}$ and red lines for $P_{2}$). In this case, if $\underline{\mu}_{1}=5<5.4=4\underline{\mu}/3$ say, then stable cavitation is expected, but there is also about 10\% chance that unstable snap cavitation occurs. Similarly, when $4\underline{\mu}/3=5.4<\underline{\mu}_{1}=5.8<8.1=2\underline{\mu}$, unstable cavitation is expected, but there is also about 10\% chance that the cavitation is stable. Stable and unstable cavitation of a stochastic sphere are illustrated numerically in Figure~\ref{fig:sphere-stdiag}. Specifically:
\begin{itemize}
\item[(a)] In Figure~\ref{fig:sphere-stdiag}(a), $b=0$ in \eqref{eq:R12:b}, and the random variable $R_1=\mu_{1}/\mu$ is drawn from a Beta distribution with $\xi_{1}=287$ and $\xi_{2}=36$. In this case, $\underline{\mu}_{1}=3.6<5.4=4\underline{\mu}/3$, and stable cavitation, with supercritical bifurcation after the spherical cavity opens, is expected.

\item[(b)] In Figure~\ref{fig:sphere-stdiag}(b), $b=-3$ in \eqref{eq:R12:b}, and the random variable $R_1=(\mu_{1}+3)/(\mu+6)$ draws its values from a Beta distribution with $\xi_{1}=325$ and $\xi_{2}=10$. Thus, $4\underline{\mu}/3=5.4<\underline{\mu}_{1}=6.75<8.1=2\underline{\mu}$, and unstable cavitation, with subcritical bifurcation after the spherical cavity forms, is expected.
\end{itemize}

For the numerical examples shown in Figure~\ref{fig:sphere-stdiag} also, the critical dead load is $P_{0}=4\mu-3\mu_{1}/2$, as given by \eqref{eq:sphere:P0:st}, with $\mu$ and $\mu_{1}$ following probability distributions. In each case, the expectation is that the onset of cavitation occurs at the mean value $\underline{P_{0}}=4\underline{\mu}-3\underline{\mu}_{1}/2$, found at the intersection of the dashed black line with the horizontal axis. However, there is a chance that cavity can form under smaller or greater critical loads that the expected load value, as shown by the coloured interval about the mean value along the horizontal axis.

To summarise, for a stochastic elastic sphere under uniform tensile dead load, we obtain the probabilities of stable or unstable cavitation, given that the material parameters are generated from known probability density functions. In the deterministic elastic case, there is a single critical parameter value that strictly separates the cases where the initiation of either stable or unstable cavitation occurs. By contrast, in the stochastic case, there is a probabilistic interval, containing the deterministic critical value, where there is always a competition between the stable and unstable states in the sense that both have a quantifiable chance to be found. For the onset of cavitation, there is also a probabilistic interval where a cavity may form, with a given probability, under smaller or greater loads that the expected critical value.

%%%%%%%%%%%%%%%%%%%%%%%%%%%%%%%%%%%%%%%%%%%%%%%%%%%%%%%%%%%%
%%%%%%%%%%%%%%%%%%%%   NEW SECTION  %%%%%%%%%%%%%%%%%%%%%%%%
%%%%%%%%%%%%%%%%%%%%%%%%%%%%%%%%%%%%%%%%%%%%%%%%%%%%%%%%%%%%
\section{Conclusion}\label{sec:conclude}

This work is motivated by the fact that a crucial part in assessing the physical properties of many solid materials is to quantify the uncertainties in their mechanical responses, which cannot be ignored. In particular, the idealised problem of the formation of a spherical cavity at the centre of a solid sphere illustrates some important effects on the likely elastic responses of stochastic hyperelastic materials under large strains.

For homogeneous isotropic incompressible spheres of stochastic hyperelastic material, subject to radial tensile dead loads applied uniformly on the sphere surface, we examined the possible radially symmetric deformations and determined which of these deformations are stable. Homogeneous stochastic hyperelastic material models satisfying certain theoretical assumptions were recently introduced to capture the dispersion in experimental data in addition to the traditional mean-data values \cite{Mihai:2018a:MWG,Staber:2017:SG}.

For the deterministic elastic problem, where the model parameters are single-valued constants, non-trivial deformations, whereby a spherical cavity forms at the centre of the sphere, are possible for some classes of materials when the applied tensile dead loads are sufficiently large \cite{Ball:1982}. In some materials, cavitation is stable, in the sense that the cavity radius monotonically increases as the applied dead load increases \cite{ChouWang:1989:CWH}. Here, we showed that a sudden jump in the cavity opening, causing unstable snap cavitation, at the critical dead load can also occur in a homogeneous isotropic incompressible sphere, provided that the material satisfies Baker-Ericksen inequalities. If such a material could be found, a sphere made of this material would suddenly increase its volume at a critical load and show some form of hysteresis as the load is removed.

In the stochastic case, the probabilistic nature of the solution reflects the probability in the constitutive law, and bifurcation and stability can be quantified in terms of probabilities. By contrast to the deterministic elastic problem, where deterministic critical parameter values strictly separate the cases where either the stable or unstable cavitation occurs, for the stochastic problem, we obtained probabilistic intervals where both states have a quantifiable chance to exist. For the onset of cavitation, there is a probabilistic interval where the cavity may form, with a given probability, under smaller or greater loads that the expected critical value.

As a direct application of our approach, one could consider the cavitation of an inhomogeneous sphere made of concentric homogeneous spheres of different stochastic material, similar to the concentric homogeneous spheres of deterministic elastic material treated in \cite{Horgan:1989:HP} and \cite{Sivaloganathan:1992}. Such composite spheres would require comparing both ensemble and spatial averages.

%%%%%%%%%%%%%%%%%%%%%%%%%%%%%%%%%%%%%%%%%%%%%%%%%%%%%%%%%%%%
%%%%%%%%%%%%%%%%%%%%   NEW SECTION  %%%%%%%%%%%%%%%%%%%%%%%%
%%%%%%%%%%%%%%%%%%%%%%%%%%%%%%%%%%%%%%%%%%%%%%%%%%%%%%%%%%%%
\appendix
\section{Stability analysis}\label{sec:append}

We provide a corrected version of Proposition 5.2 of \cite{Ball:1982} and its proof. In particular, we show that, in the deterministic elastic case, both subcritical and supercritical behaviours close to the bifurcation are possible, depending on the material.

\begin{proposition}\label{app:prop}
Let $W(\lambda)$ be twice differentiable at $\lambda=1$, and
\[
P(c)=\left(1+c^{n}\right)^{(n-1)/n}\int_{\left(1+c^{n}\right)^{1/n}}^{\infty}\frac{\text{d}W}{\text{d}\lambda}\frac{\text{d}\lambda}{\lambda^{n}-1},
\]
where $n>1$. Then $\lim_{c\to0_{+}}\left(\text{d}P/\text{d}c\right)=0$, and if
\begin{equation}\label{eq:prop1}
\lim_{c\to0_{+}}\left(P(c)-\lim_{\lambda\to1}\frac{1}{n(n-1)}\frac{\text{d}^2W}{\text{d}\lambda^2}\right)>0,
\end{equation}
then $\text{d}P/\text{d}c>0$ for sufficiently small $c>0$ (i.e., the bifurcation is supercritical), while if
\begin{equation}\label{eq:prop2}
\lim_{c\to0_{+}}\left(P(c)-\lim_{\lambda\to1}\frac{1}{n(n-1)}\frac{\text{d}^2W}{\text{d}\lambda^2}\right)<0,
\end{equation}
then $\text{d}P/\text{d}c<0$ for sufficiently small $c>0$ (i.e., the bifurcation is subcritical).
\end{proposition}
These cases are illustrated, for the particular example of material presented in this paper, in Figures~\ref{fig:subcav} and~\ref{fig:bif}.

\paragraph{Proof.} We denote $\theta=\left(1+c^{n}\right)^{(n-1)/n}$ and define $\widehat{P}(\theta)=P(c)$. Then
\[
\widehat{P}(\theta)=\theta\int_{\theta^{1/(n-1)}}^{\infty}\frac{\text{d}W}{\text{d}\lambda}\frac{\text{d}\lambda}{\lambda^{n}-1}
\]
and
\[
\frac{\text{d}P}{\text{d}c}=\frac{\text{d}\widehat{P}}{\text{d}\theta}\frac{\text{d}\theta}{\text{d}c},
\]
where
\[
\frac{\text{d}\widehat{P}}{\text{d}\theta}=\int_{\theta^{1/(n-1)}}^{\infty}\frac{\text{d}W}{\text{d}\lambda}\frac{\text{d}\lambda}{\lambda^{n}-1}-\frac{\theta^{1/(n-1)}}{n-1}\left(\frac{\text{d}W}{\text{d}\lambda}\frac{1}{\lambda^{n}-1}\right)\left|_{\lambda=\theta^{1/(n-1)}}\right..
\]
It follows that
\[
\begin{split}
\lim_{\theta\to1}\frac{\text{d}\widehat{P}}{\text{d}\theta}&=\lim_{\theta\to1}\int_{\theta^{1/(n-1)}}^{\infty}\frac{\text{d}W}{\text{d}\lambda}\frac{\text{d}\lambda}{\lambda^{n}-1}-\lim_{\theta\to1}\frac{\theta^{1/(n-1)}}{n-1}\frac{\text{d}W}{\text{d}\lambda}\frac{1}{\lambda^{n}-1}\left|_{\lambda=\theta^{1/(n-1)}}\right.\\
&=\lim_{c\to0_{+}}P(c)-\lim_{\theta\to1}\frac{\theta^{1/(n-1)}}{n-1}\left(\frac{\text{d}W}{\text{d}\lambda}\frac{1}{\lambda^{n}-1}\right)\left|_{\lambda=\theta^{1/(n-1)}}\right.\\
&=\lim_{c\to0_{+}}P(c)-\lim_{\lambda\to1}\frac{1}{n(n-1)}\frac{\text{d}W}{\text{d}\lambda}\frac{1}{\lambda-1}\\
&=\lim_{c\to0_{+}}P(c)-\lim_{\lambda\to1}\frac{1}{n(n-1)}\frac{\text{d}^2W}{\text{d}\lambda^2}.
\end{split}
\]
Hence,
\begin{equation}\label{eq:final}
\lim_{c\to0_{+}}\frac{\text{d}P}{\text{d}c}=\lim_{\theta\to1}\frac{\text{d}\widehat{P}}{\text{d}\theta}=\lim_{c\to0_{+}}P(c)-\lim_{\lambda\to1}\frac{1}{n(n-1)}\frac{\text{d}^2W}{\text{d}\lambda^2}
\end{equation}
and $\text{d}P/\text{d}c>0$ (respectively, $\text{d}P/\text{d}c<0$) for sufficiently small $c>0$ if and only if \eqref{eq:prop1} (respectively, \eqref{eq:prop2}) holds. This concludes the proof.\\

Note that the difference between this result and Proposition 5.2 of \cite{Ball:1982} comes from the (correct) minus sign between the two terms on the right-hand side of~\eqref{eq:final} (whereas a plus sign is found in the corresponding unlabelled expression appearing between Equations (5.25) and (5.26) of \cite{Ball:1982}).

%%%%%%%%%%%%%%%%%%%%%%%%%%%%%%%%%%%

%\paragraph{Data access.} There are no supplementary data associated with this paper.

%\paragraph{Authors contribution.} All authors contributed equally to all aspects of this article and gave final approval for publication.

%\paragraph{Competing interests.} The authors declare that they have no competing interests.

\paragraph{Acknowledgement.} We thank  John Ball for a discussion on the corrected version of Proposition 5.2 of \cite{Ball:1982}, as presented here in Appendix~\ref{sec:append}. The support for Alain Goriely by the Engineering and Physical Sciences Research Council of Great Britain under research grant EP/R020205/1 is gratefully acknowledged.

%\paragraph{Funding.} The support for Alain Goriely by the Engineering and Physical Sciences Research Council of Great Britain under research grant EP/R020205/1 is gratefully acknowledged.

%%%%%%%%%%%%%%%%%%%%%%%%%%%%%

%%%%%%%%%%%%%%%%%%%
\end{document}